\documentclass[reprint,amsmath,amssymb,aps,prl]{revtex4-1}

\usepackage{graphicx,subfigure,bm,amssymb,amsmath,dcolumn,hyperref}
\usepackage{color,multirow}
\usepackage{mathrsfs}

\begin{document}
\definecolor{blue}{rgb}{0,0,1}
\definecolor{red}{rgb}{1,0,0}
\newcommand{\blue}[1]{\textcolor{blue}{#1}}
\newcommand{\red}[1]{\textcolor{red}{#1}}

\newcommand{\Kxc}{K_{\rm xc}}
\newcommand{\Tf}{T_{\textsc f}}
\newcommand{\kf}{k_{\textsc f}}
\newcommand{\qm}{q_{\rm max}}
\newcommand{\Ek}{E_{\rm k}}
\newcommand{\Eko}{E_{\rm k0}}

\preprint{APS/123-QED}

\title{Exchange-Correlation Effect in the Charge Response of a Warm Dense Electron Gas}

\author{Peng-Cheng Hou$^{1}$}
\author{Bao-Zong Wang$^{1}$} 

\author{Kristjan Haule$^{4}$}

\author{Youjin Deng$^{1,2,3}$}
\author{Kun Chen$^{5}$}
\email{kunchen@flatironinstitute.org}

\affiliation{$^{1}$ Department of Modern Physics, University of Science and Technology of China, Hefei, Anhui 230026, China}
\affiliation{$^{2}$ Hefei National Laboratory, University of Science and Technology of China, Hefei 230088, China}
\affiliation{$^{3}$ MinJiang Collaborative Center for Theoretical Physics,
College of Physics and Electronic Information Engineering, Minjiang University, Fuzhou 350108, China}

\affiliation{$^{4}$ Department of Physics and Astronomy, Rutgers,
The State University of New Jersey, Piscataway, New Jersey 08854-8019 USA}

\affiliation{$^{5}$ Center for Computational Quantum Physics, Flatiron Institute, 162 5th Avenue, New York, New York 10010. The Flatiron Institute is a division of the Simons Foundation}

\date{\today}

\begin{abstract}
The study of warm dense matter, widely existing in nature and laboratories,
is challenging due to the interplay of quantum and classical fluctuations. 
We develop a variational diagrammatic Monte Carlo method and determine the exchange-correlation kernel $\Kxc(q;T)$
of a warm dense electron gas for a broad range of temperature $T$ and momentum $q$. We observe several interesting physics, including the $T$-dependent evolution of the hump structure and the large-$q$ tail and the emergence of a scaling relation.
Particularly, by deriving an analytical form for $q \to \infty$, we obtain large-$q$ tails of $\Kxc$ with high precision. It is shown that the $\Kxc$  data can be reliably transformed into real space, which can be directly used in density-functional-theory calculations of real warm dense matter.
\end{abstract}

\maketitle

Over the past two decades, extensive research interests have been devoted to an extreme state of matter at high temperature and density,
called warm dense matter (WDM). The WDM region occurs in various astrophysical objects such as giant planet 
interiors~\cite{saumon_shock_2004,vorberger_hydrogen-helium_2007,nettelmann_ab_2008,militzer_massive_2008,schottler_ab_2018}, 
brown dwarfs~\cite{saumon_role_1992,becker_ab_2014}, and neutron crusts~\cite{daligault_electronion_2009}. 
Nowadays, it can be routinely realized and probed in novel high-power laser and accelerator 
facilities~\cite{zastrau_resolving_2014, fletcher_ultrabright_2015,sperling_free-electron_2015, tschentscher_photon_2017}. 
The electrons in WDM are about to lose quantum coherence while still demonstrating nontrivial quantum corrections, 
resulting in a special crossover state between classical plasma and quantum condensed matter. 

Within the local-density approximation treatment, the electrons in WDM are modeled by the warm dense uniform electron gas (UEG).
The UEG only has two parameters, the density parameter (Wigner-Seitz radius) $r_s=\Bar{r}/a_{\textsc b}$ and the reduced temperature $\theta=T/\Tf$, 
where $\Bar{r}$ is the average interparticle distance, $a_{\textsc b}$ is Bohr radius, and $\Tf$ is the Fermi temperature. At the high-density limit ($r_s\to 0 $), the electrons behave as an ideal Fermi gas; at low density, the Coulomb potential becomes dominant, eventually resulting in a Wigner crystal~\cite{wigner_interaction_1934,wigner1938effects,drummond_diffusion_2004}.
Dimensional temperature ratio $\theta$ can be seen as a quantum degeneracy parameter. In the high-temperature ($\theta \gg 1$) and zero-temperature ($\theta \to 0$) limits, the UEG has been well described by the classical plasma theory and the quantum condensed matter theory, respectively.
However, due to the complicated interplay of electronic correlations, quantum coherence, and thermal fluctuations, 
the crossover regime $r_s \sim \theta \sim 1$, particularly relevant for WDM, is much less understood. 

The density (or charge) response function, one of the most important electronic probes, is measured in many vital diagnostic experiments of WDM~\cite{glenzer_x-ray_2009,baczewski_x-ray_2016,Kraus-characterizing_2018,white_observation_2012,denoeud_dynamic_2016,medvedev_multistep_2018}. The random phase approximation (RPA) well describes the long-range screening effects and is a reasonable approximation in the weakly interacting regimes. However, 
it becomes invalid in the WDM conditions because the RPA response function only accounts for the Hartree component of the induced potentials~\cite{olsen_beyond_2019} and overestimates the short-range correlations between electrons. The neglected exchange-correlation (XC) component can be parameterized by the local field correction (LFC),
\begin{equation}
    G(q,\omega)= 1- \frac{1}{v_q}\left[\frac{1}{\chi_0(q,\omega)}-\frac{1}{\chi(q, \omega)}\right] ,
\label{eq:LFC}
\end{equation}
where $\chi_0(q,\omega)$ and $\chi(q,\omega)$ are the dynamic density response functions of noninteracting and interacting systems, respectively, 
and $v_q=8\pi/q^2$ is the long-range Coulomb repulsion (Rydberg atomic units are used).
The LFC encodes the structure of effective electron-electron interaction and is crucial for the understanding of many experimental phenomena, 
exemplified by plasmon~\cite{glenzer_observations_2007,sperling_free-electron_2015}, 
Coulomb and spin-Coulomb drag effects~\cite{badalyan_finite_2008,upadhyay_study_2021}, 
electrical and thermal conductivities~\cite{desjarlais_density-functional_2017,veysman_optical_2016}, 
stopping power~\cite{cayzac_experimental_2017,fu_energy_2017}, and energy transfer rates~\cite{PhysRevE.81.046404}. 
In the time-dependent density functional theory (TDDFT)~\cite{runge_density-functional_1984,ullrich_time-dependent_2012,olsen_beyond_2019}, 
the so-called XC kernel, $\Kxc(q, \omega)=-v_q G(q, \omega)$, is typically used as an input for \emph{ab-inito} calculations of ground-state energies and electronic spectra in real materials. 
For small- and large-$q$, since $G(q) \propto q^2$, the behavior of $G$ is better described by $\Kxc$.

Substantial efforts have been devoted to studying the LFC of UEG, particularly its static one $G(q) \equiv G(q,\omega=0)$.
At $T=0$, in the ground state, analytic properties of the LFC in the small- and large-$q$ limits were derived~\cite{gross_local_1985,iwamoto_correlation_1987,rogers_exact_1987,vignale_exact_1988} and numerical calculations of $G(q)$ were performed for various density parameters by diffusion quantum Monte Carlo~\cite{bowen_static_1994,moroni_static_1995}. 
Recently, Chen and Haule developed a variational diagrammatic Monte Carlo (VDMC) method~\cite{chen_combined_2019} using renormalized Feynman diagrammatic expansion and used it to obtain accurate results on the static density LFC~\cite{chen_combined_2019} and spin LFC~\cite{kukkonen_quantitative_2021} for $r_s \leq 5$. Furthermore, the dynamic LFC was computed by implementing the algorithmic Matsubara integration with the VDMC~\cite{leblanc_dynamic_2022}.
For finite $T$, calculations of $G(q)$ are amenable due to the recent development of unbiased path-integral Monte Carlo (PIMC) methods~\cite{dornheim_permutation_2015,doi:10.1063/1.4936145,schoof_ab_2015,groth_ab_2016}. 
Dornheim \textit{et al.}~\cite{dornheim_static_2019} reported a neural-network representation of $G(q)$ 
based on PIMC data and ground-state parametrization~\cite{corradini_analytical_1998}.
They further gave an analytic parametrization of $G(q)$ within effective static approximation~\cite{dornheim_effective_2020} and used it to describe the response functions in the WDM regime~\cite{dornheim_analytical_2021,dornheim_density_2021}.

Many challenges still exist. First, due to the limited precision of ground-state data, 
it remains controversial whether the LFC develops a peak structure near $q=2\kf$ (where $\kf$ is the Fermi momentum), 
termed the ``$2\kf$ hump puzzle''~\cite{giuliani_vignale_2005}. 
Second, the PIMC methods suffer from the notorious sign problem~\cite{troyer_computational_2005,dornheim_path_2019,dornheim_fermion_2019} as the electrons become a degenerate quantum liquid for $T<\Tf$~\cite{dornheim_uniform_2018}. 
Therefore, our current knowledge has a significant gap: how the LFC evolves from zero $T$ in the ground state, via low $T$ relevant for condensed matter physics and intermediate $T$ corresponding to WDM, and finally to high $T$ relevant for classical plasma. 
Third, the PIMC extracts $G(q)$ from $\chi(q)$ via~\eqref{eq:LFC}, but it is difficult to determine the large-$q$ LFC because statistical errors in MC data can easily overwhelm the very weak many-body contribution to $\chi(q)-\chi_0(q)$ for $q> 3\kf$. 
To develop an efficient algorithm for large $q$, it is desirable to analytically derive the asymptotic properties of $G(q)$ in the large-$q$ limit. 
In application, this is particularly important since many TDDFT algorithms require an XC kernel $\Kxc(r)$ in real space, and, 
without accurate data for large $q$ and analytical forms for $q\to \infty$, a reliable Fourier transform from $\Kxc(q)$ to $\Kxc(r)$ is extremely difficult due to the Gibbs phenomenon.

\begin{figure}
    \centering
    \includegraphics[width=0.70\linewidth]{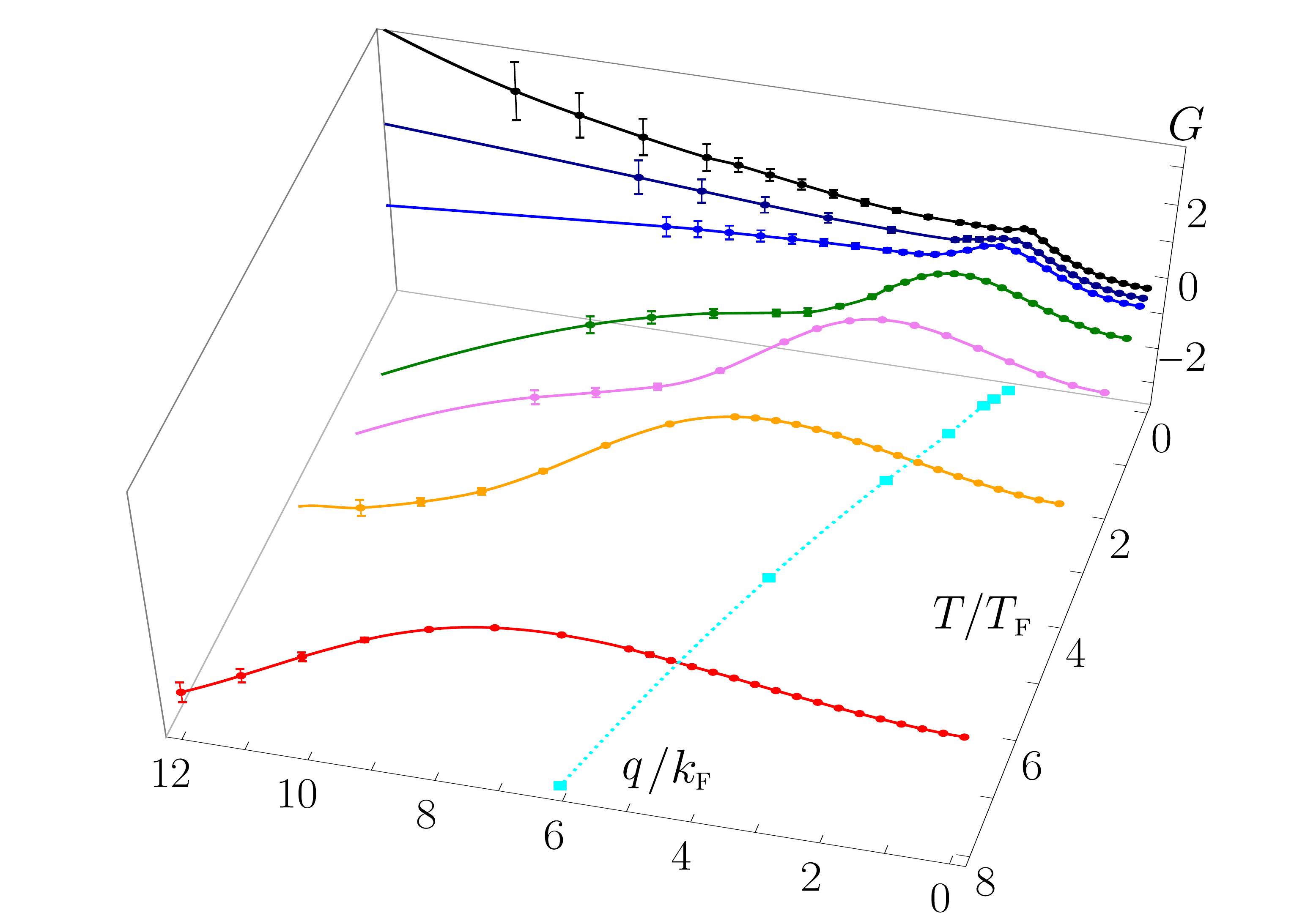}
    \caption{Static local field correction $G(q)$ in the extensive ($q,T$) plane for $r_s=1$. The circles depict VDMC data points for $\theta=0.0625, 0.25, 0.4, 1, 2, 4, 8$. The cyan squares and dotted line depict the tendency of the maximum momentum $\qm$ of $G(q)$ versus $T$.}
	    \label{fig:LFC_qT}
\end{figure}

\begin{figure*}
    \centering
    \includegraphics[width=0.95\linewidth]{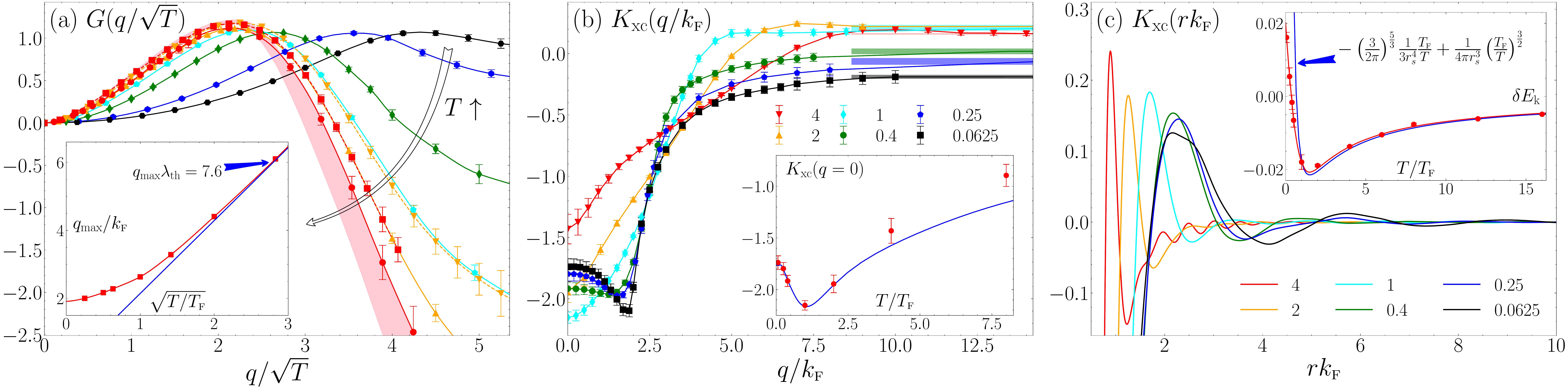}
    \caption{(a) Rescaled static LFC as a function of $q/\sqrt{T}$. The data points with solid lines are for $r_s=1$ and $\theta=0.25, 0.4,1,2,4, 8$, and ones with dashed lines are for $r_s=2$ and $\theta=4,8$. As $T\to \infty$, the LFC data tend to collapse to a universal curve that depends on $q/\sqrt T$, and is covered in the pink shadow. The inset shows the evolution of $\qm$ from $\simeq 2\kf$ in the zero-$T$ limit to $\approx 7.6 \lambda_{\rm th}$ in the high-$T$ limit. 
    (b) The XC kernel $\Kxc(q)$ for different temperatures and $r_s=1$. The tail of either curve converges to a constant $K_{\infty}$ that is explicitly calculated by the VDMC as Eq.\eqref{eq:LFC_Tlq}, and the strip marks its value and error. The inset shows the evolution of $\Kxc(q=0,T)$. Our data are consistent with the parameterized function~\cite{groth_ab_2017} (blue line).
    (c) The XC kernel of UEG in real space with the delta-peaked term subtracted. The inset shows the interaction-induced shift of the kinetic energy versus temperature. Our data is approaching the high-$T$ approximation~\cite{kraeft_kinetic_2002} (blue line) for $\theta > 1$.}
	    \label{fig:mainresults}
\end{figure*}

In this letter, we develop the VDMC method for the warm dense electron gas, which is a generalization of the method
introduced by two of us~\cite{chen_combined_2019}. 
Inspired by the binary expression tree in computer science, we introduce an expression tree representation of Feynman diagrams 
to efficiently and accurately calculate $\chi(q)$.
Our VDMC method is reliable for the UEG of high to intermediate density and at arbitrary temperature, 
covering those low-$T$ and large-$q$ regimes where other unbiased finite-$T$ quantum MC methods are infeasible.
From extensive simulations, we obtain high-precision data of the static LFC in a broad range of $(q,T)$ space 
from the Fermi liquid to WDM and then to classical plasma. Several interesting physics are revealed, including the $T$-dependent evolution of the hump structure and the large-$q$ tail (Fig.~\ref{fig:LFC_qT}) and the emergence of an asymptotic high-$T$ scaling relation of $G(q;r_s,T)$ (we use a semicolon to separate intrinsic variables including $q$, $\omega$, and $r$, and parameter variables including $r_s$ and $T$) in dimensionless variable $q/\sqrt{T}$ (Fig.~\ref{fig:mainresults}(a)). High-precision data for $\Kxc$ are obtained up to $14\kf$, shown in Fig.~\ref{fig:mainresults}(b).
In both the small- and the large-$q$ limit,  $\Kxc(q)$ saturates to a constant, non-monotonically depending on $T$.
Moreover, we show that the asymptotic XC kernel $K_\infty \equiv \Kxc(q \! \to \! \infty)$ is proportional to the interaction-induced shift of the kinetic energy and derive the analytical form of the coefficient.  
Using a VDMC, we compute the kinetic energy accurately and thus $K_\infty$ as a function of $T$ (the inset of Fig. 2(c)).
Combining data of $\Kxc(q)$ for $q \leq 14 \kf$ and of $K_\infty$ for the tail, 
we show that the Fourier transform of $\Kxc(q)$ to real space can be performed reliably without any freely adjustable parameter (Fig.~\ref{fig:mainresults}(c)). 

{\em{Results -}} We carry out extensive simulations for the three-dimensional UEG model at a series of temperatures $\theta=0.0625, 0.25, 0.4, 1, 2, 4, 8$. Hereby, we focus on parameter $r_s=1$, which is the most typical WDM density, and we also simulate $r_s=2$ for $\theta=4,8$ to illustrate the $r_s$-independent scaling relation.
We accurately determine the static LFC $G(q;T)$ up to $q/\kf=14$ (Fig.~\ref{fig:LFC_qT}), more than four times larger than the previous studies~\cite{groth_ab_2019,dornheim_static_2019}. Accordingly, the XC kernel $\Kxc(q;T)$ is obtained in Fig.~\ref{fig:mainresults}(b).

Our data exhibit several profound features of the $T$- and $q$-resolved LFC.
For any given $T$, there is a unique momentum $\qm$ corresponding to the local peak in $G(q)$, which separates long-range (small-$q$) from short-range (large-$q$) physics. Figure~\ref{fig:LFC_qT} shows that the peak broadens and shifts to the larger momentum as $T$ increases, and the inset of Fig.~\ref{fig:mainresults}(a) shows $\qm$ as a function of $T$. In the high-$T$ limit, $\qm$ is controlled by the thermal de Broglie wavelength $\lambda_{\rm th}=\sqrt{4\pi/T}$ as $\qm \approx 7.6\lambda_{\rm th}^{-1}$. As $T$ decreases, the value of $\qm$ gradually decreases and is no longer dominated by $\lambda_{\rm th}$ for $T \lesssim 2\Tf$, eventually converging to $\simeq 2\kf$ at zero $T$.
We further observe that in the evolution of the XC kernel $\Kxc(q)$ with $T$ (Fig.~\ref{fig:mainresults}(b)), a local minimum appears at $T \lesssim 0.4 \Tf$ and becomes sharper and closer to $q \simeq 2\kf$ as $T\to 0$. These numerical results confirm the ``$2\kf$ hump" of the ground-state UEG postulated in various theoretical predictions~\cite{geldart_wave-number_1970,brosens_dielectric_1980,farid_extremal_1993,richardson_dynamical_1994}. Moreover, the nontrivial $T$-dependence of $\qm$ and the smearing out of the sharp minimum in $\Kxc$ at $T\sim\Tf$ reflect the intense competition between quantum and thermal fluctuations in the WDM regime.

In addition, with increasing $T$, the large-$q$ LFC evolves from a positive tail at low $T$ to a negative tail for $T\gtrsim 0.4 \Tf$. Holas~\cite{rogers_exact_1987} predicted the asymptotic behavior $G(q)\propto q^2$ for $q\to \infty$ at zero $T$, where the prefactor is proportional to the change in the interaction-induced shift of the kinetic energy $\delta \Ek$ and is positive for all $r_s$. However, no analytical formula currently exists for finite $T$. Qualitatively, the evolution of the LFC tail reflects the sophisticated competition of Coulomb repulsion, exchange effects, and thermal motion.

We argue that in the high-$T$ limit, the entire LFC curve is described by an asymptotic scaling function depending on a single dimensionless parameter $q/\sqrt T$, independent of $r_s$, as 
\begin{equation}
    \lim_{T \to \infty}\rvert_{{\rm fixed}\; \alpha } G(q;r_s, T)=\tilde G(\alpha) \quad \alpha=q/\sqrt{T} \;.
\label{eq:scaling}
\end{equation}
The argument is as follows: for UEG, there are two characteristic length scales, i.e., $\lambda_{\rm th}$ and $r_s$.
When $T\to \infty$, the only detectable length scale is $\lambda_{\rm th}$ since $\lambda_{\rm th} \ll r_s$. 
Indeed, Figure~\ref{fig:mainresults}(a) demonstrates that as $T$ increases, the LFC tends to collapse into a universal curve as Eq.~\ref{eq:scaling}.

We now turn to the XC kernel.
Figure~\ref{fig:mainresults}(b) shows that $\Kxc(q)$ converges to a nonmonotonic $T$-dependent constant for both $q\to 0$ and $q\to \infty$. 
It is known that the uniform XC kernel $\Kxc(q=0)$ is proportional to the second derivative of the XC free energy, 
given by the compressibility sum rule~\cite{martin_sum_1988}. Therefore, we compare our $\Kxc(q=0)$ data with the values from the recent XC free-energy parametrization~\cite{groth_ab_2017} in the inset of Fig.~\ref{fig:mainresults}(b). It is shown that they are consistent for $\theta \leq 2$ within error bars while having a slight deviation for $\theta \geq 4$. 

The large-$q$ behavior of the XC kernel is highly nontrivial, as also discussed in the LFC tail. Based on the large-$q$ expansion of the polarization function $\Pi$ related to the density response function by $\chi(q,\omega)= \left[\Pi^{-1}(q,\omega)+v_q \right]^{-1}$ (details are given in the Supplementary Material~\cite{supplement}), we derive the asymptotic formula of the large-$q$ XC kernel for finite $T$ as
\begin{equation}
    \lim_{q\to \infty} \Kxc(q;r_s, T)= K_{\infty}(r_s,T)+ \mathcal O(q^{-2}) \;, 
  \label{eq:LFC_Tlq}
\end{equation}
with $ K_{\infty} = -(32\pi^2 r_s^6 /27) \delta \Ek$.
This analytic formula generalizes the zero-$T$ one~\cite{rogers_exact_1987} and describes that the convergent constant $K_{\infty}$ is controlled by the many-body contributions of the kinetic energy. Note that Eq.~\ref{eq:LFC_Tlq} still holds for the dynamic XC kernel since the frequency dependence only exists in the subleading and higher orders. We explicitly calculate the kinetic energies by the VDMC to extract $K_{\infty}$ for various temperatures, and the results are consistent with our large-$q$ $\Kxc$ data within error bars, as shown in Fig.~\ref{fig:mainresults}(b). Furthermore, the inset of Fig.~\ref{fig:mainresults}(c) displays an interesting nonmonotonic behavior: as $T$ increases, $\delta \Ek$ decreases from a positive value for low $T$ to negative for $\theta \gtrsim 0.4$ and finally approaches zero asymptotically.
At zero $T$, $\delta \Ek$ is always positive because electron repulsion broadens the ground-state momentum distribution. However, at high enough temperatures, the correlation effects would lead to the narrowing of the momentum distribution~\cite{militzer_lowering_2002,kraeft_kinetic_2002,hunger_momentum_2021}; meanwhile, the Hartree-Fock term $-(3/2\pi)^{5/3}\theta/(3r_s^4)$ dominates the XC contribution to the kinetic energy so that $\delta \Ek<0$. Remarkably, we find that the high-$T$ approximation~\cite{kraeft_kinetic_2002}, including the Hartree-Fock and the Montroll-Ward contributions, describes well the nonmonotonic $T$-dependence of $\delta \Ek$ for $\theta \geq 1$. 
The formula~\eqref{eq:LFC_Tlq} combined with the behavior of $\delta \Ek$ versus $T$ explains the $T$-dependent evolution of the LFC/XC kernel tail and matches the scaling relation in the high-$T$ limit.

Combining our numerical data and the analytical tail, we perform interpolation and obtain a controlled representation of the XC kernel for overall momentum (the solid line of Fig.~\ref{fig:mainresults}(b)). In practice, we use spline interpolation to fit piecewise cubic polynomials to $q \leq q_{\rm cut}$ data (the cutoff momentum $q_{\rm cut}$ excludes the tail and depends on the temperature). We then perform the least-squares fits to $q > q_{\rm cut}$ data via the ansatz $K_{\rm xc}(q) = K_{\infty} + a /q^2$, where $K_{\infty}$ is fixed by $\delta E_{\rm k}$ data and $a$ is the fitting parameter. The overall $\Kxc(q)$ enables a Fourier transform to the real-space XC kernel $\Kxc(r)$ without relying on any ansatz of $\Kxc(q)$. The constant tail in $\Kxc(q)$ translates to a contact term $K_{\infty}\delta^{(3)}(r)$ in real space, while the rest part of the XC kernel transforms into a smooth function of the distance $r$, as shown in Fig.~\ref{fig:mainresults}(c). This scheme allows us to accurately represent $\Kxc(r)$, which is a key input in TDDFT calculations for real materials.

{\em{Methods -}}
Diagrammatic expansion is used to perturbatively calculate a quantity as a series of integrals which can be visualized as Feynman diagrams. 
It allows us to simulate quantities immediately in the thermodynamic limit by a Markov-chain process that stochastically samples diagrams and internal variables.
Such diagrammatic MC methods have found successful applications in many physical problems~\cite{PhysRevLett.81.2514,Kozik_2010,PhysRevLett.113.195301,deng2015emergent,PhysRevB.96.041105,PhysRevB.96.081117,PhysRevB.97.085117,PhysRevB.99.035120,PhysRevB.100.121102,PhysRevB.101.075113,PhysRevLett.124.117602,van2012feynman,PhysRevB.99.035140,PhysRevLett.121.130405,PhysRevLett.121.130406,PhysRevB.62.6317,Prokof2008Fermi,PhysRevB.77.125101,PhysRevB.90.104510,PhysRevLett.113.166402,PhysRevB.97.134305,PhysRevLett.123.076601,PhysRevB.101.045134,PhysRevLett.110.070601,PhysRevB.87.024407,PhysRevLett.116.177203}. 
For UEG, the expansion in the bare Coulomb interaction is divergent and needs to be transformed into a more appropriate power expansion, as explained below. We use the variational scheme~\cite{chen_combined_2019,haule_single-particle_2022}, 
in which the emergent low-energy physics is taken into account at the lowest order, and the corrections are perturbatively added, leading to rapid convergence. We develop a VDMC method, which has a generic algorithmic structure and an optimized efficiency with the help of an expression-tree representation of diagrams. 

\begin{figure}
\begin{center}
    \includegraphics[width=\linewidth]{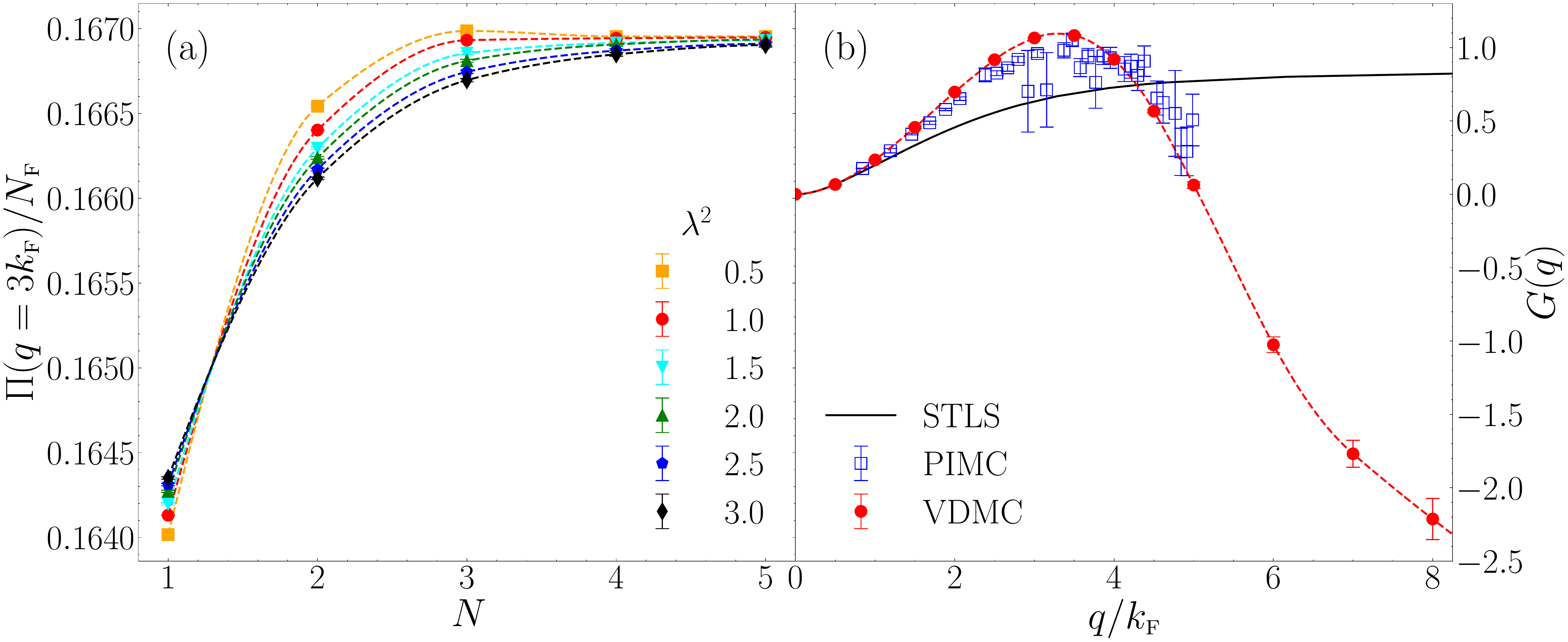}
    \caption{    
    (a) Static polarization scaled by the density of states at the Fermi level $N_{\textsc f}$ at $q=3\kf$ versus truncation order $N$ for $\theta=2$ and $r_s=1$. All $\lambda$ choices lead to the same extrapolated value, and the optimal $\lambda$ for the fastest convergence is about $1.0$ in this case.  (b) LFC $G(q)$ at $\theta=2$ and $r_s=1$ for the VDMC (red circles), the PIMC~\cite{dornheim_static_2019}(blue squares), and the STLS approximation~\cite{tanaka_thermodynamics_1986} (black line).
    }
    \label{fig:algorithm}
\end{center}
\end{figure}

Motivated by the Coulomb screening effects, we introduce a variational inverse screening length $\lambda_q$ following Ref.~\cite{chen_combined_2019} to re-expand the bare interaction $\frac{8\pi }{k^2} = \frac{8\pi }{k^2+\lambda_q^2} \sum_{n=0}^{\infty} \left(\frac{\lambda_q^2}{8\pi} \frac{8\pi}{k^2+\lambda_q^2}\right)^n$. 
In the VDMC, $\lambda_q$ is $q$-dependent and optimized to achieve a rapid convergence according to the principle of minimal sensitivity~\cite{kleinert_path_1995}. 
To further speed up the convergence, we insert the Fock sub-diagram into the bare electron propagator to recover the screened Hartree-Fock approximation in the first order. Meanwhile, we add order-by-order chemical-potential counterterms to fix the Fermi surface at each order.
 Within this optimized expansion, we can obtain reliable infinite-order results of arbitrary quantities without very large truncation order $N$, which avoids the exponential scaling of the computation time with order. Fig.~\ref{fig:algorithm}(a) shows the static polarization $\Pi(q)$, which rapidly converges at the optimal $\lambda_q$. In practical simulations, we choose $\lambda_q$ as piecewise constants for $q\lesssim 6\kf$ and $q \gtrsim 6\kf$, which is already sufficient for rapid convergence. We compute the Feynman diagrams up to order $N=5$, which contain about 1200 diagrams. Our calculations suggest that, at least for UEG, $N=5$ is already sufficient for us to reliably extrapolate $\Pi(q)$ to infinite order. The description of details, including the topology and number of Feynman diagrams and numerical approach, can be found in Ref.~\cite{chen_combined_2019}.  

A major challenge of the VDMC is how to generate and group a large number of Feynman diagrams and efficiently compute their weights.
By sign-structure analysis of scattering amplitudes in diagrams constrained by the crossing symmetry and the conservation law~\cite{wang_fermionic_2021}, we can optimize the internal-variable arrangements and construct the sign-canceled diagram groups to alleviate the sign problem significantly. Inspired by the binary expression tree in data structures, we develop a universal expression tree representation of diagrams implemented by a diagram mini-compiler with a three-layer infrastructure.
We are improving the original code for readability and usability and will report details elsewhere.
 
By the VDMC method, we obtain more accurate LFC data over a wider range of $(q,T)$ than the state-of-the-art PIMC~\cite{dornheim_static_2019}, as shown in Fig.~\ref{fig:algorithm}(b). We note that for the LFC at large momenta, the PIMC cannot provide reliable results because the statistical errors of data of $\chi$ ($\sim 1/q^2$) overwhelm $\chi-\chi_0$ ($\sim 1/q^4$)~\cite{supplement}. Here, based on the large-$q$ expansion of $\chi(q)$~\cite{supplement}, we identify each polarization diagram with a tail-decaying behavior $q^{-\ell}$, where $\ell$ is the least number of the propagator lines with flowing $q$. Hence, we optimally organize the external momentum variable of each diagram to make the integrand have the same tail behavior as the integral so that the VDMC can directly calculate $\chi-\chi_0$ by efficiently sampling each diagram with a reweighting factor $q^\ell$. For completeness, we mention that a VDMC simulation for a single $r_s-\theta$ system parameter takes O$(10^4)$ CPU (single-threaded process) hours.

{\em{Discussion -}}
In summary, we present a systematic and generic VDMC approach for warm dense electron gas and obtain high-precision results for the LFC/XC kernel over a broad range of momentum and temperature. The VDMC allows for accurate and efficient calculation of various quantities directly in the thermodynamic limit at arbitrary temperature, beyond the reach of state-of-the-art path-integral-based MC methods. Our VDMC calculations have immediate practical applications in exploring various physical systems. For example, we obtain the first \textit{ab initio} result for the real-space XC kernel, which enables finite-$T$ TDDFT calculations for real electronic systems.
A natural continuation of this work is to study the spin-dependent correlations and thermodynamic functions and construct the frequency-dependent LFC.

\begin{acknowledgments}
	 P.-C. Hou, B.-Z. Wang, and Y. Deng were supported by the National Natural Science Foundation of China (under Grant No. 11625522), the Innovation Program for Quantum Science and Technology (under grant no. 2021ZD0301900), and the National Key R\&D Program of China (under Grants No. 2018YFA0306501). K. Chen and K. Haule were supported by the Simons Collaboration on the Many Electron Problem, and K. Haule was supported by NSF DMR-1709229.
\end{acknowledgments}

\bibliographystyle{apsrev}
\bibliography{references}

\end{document}


\renewcommand{\thesection}{S-\arabic{section}}
\setcounter{section}{0}  
\renewcommand{\theequation}{S\arabic{equation}}
\setcounter{equation}{0}  
\renewcommand{\thefigure}{S\arabic{figure}}
\setcounter{figure}{0}  
\newcommand{\diff}{\mathrm{d}}
\newcommand{\biaoti}{\fontsize{12pt}{\baselineskip}\selectfont}
\newcommand{\Kxc}{K_{\rm xc}}
\newcommand{\Tf}{T_{\textsc f}}
\newcommand{\kf}{k_{\textsc f}}
\newcommand{\qm}{q_{\rm max}}
\newcommand{\Ek}{E_{\rm k}}
\newcommand{\Eko}{E_{\rm k0}}

\newcommand{\red}[1]{\textcolor{red}{#1}}

\title{\underline{Supplemental Material:} Exchange-Correlation Effect in the Charge Response of a Warm Dense Electron Gas}

\author{Peng-Cheng Hou$^{1}$}
\author{Bao-Zong Wang$^{1}$} 
\author{Kristjan Haule$^{4}$}
\author{Youjin Deng$^{1,2,3}$}
\author{Kun Chen$^{5}$}
\email{kunchen@flatironinstitute.org}

\affiliation{$^{1}$ Department of Modern Physics, University of Science and Technology of China, Hefei, Anhui 230026, China}
\affiliation{$^{2}$ Hefei National Laboratory, University of Science and Technology of China, Hefei 230088, China}
\affiliation{$^{3}$ MinJiang Collaborative Center for Theoretical Physics,
College of Physics and Electronic Information Engineering, Minjiang University, Fuzhou 350108, China}
\affiliation{$^{4}$ Department of Physics and Astronomy, Rutgers,
The State University of New Jersey, Piscataway, New Jersey 08854-8019 USA}
\affiliation{$^{5}$ Center for Computational Quantum Physics, Flatiron Institute, 162 5th Avenue, New York, New York 10010. The Flatiron Institute is a division of the Simons Foundation}

\maketitle

\section{Derivation of the asymptotic formula of the static exchange-correlation kernel in large-momentum limit}

The dynamic exchange-correlation (XC) kernel $\Kxc(q, i\nu_m)$ ($\nu_m=2m\pi/\beta$ is the bosonic Matsubara frequency) is defined as
\begin{equation}
    \Kxc(q, i\nu_m) = \frac{1}{\chi_0(q, i\nu_m)} - \frac{1}{\Pi(q, i\nu_m)} \,,
\label{eq:GPi}
\end{equation}
where $\Pi$ is the polarization function and the noninteracting density response function $\chi_0$ is actually the zero-order diagram in $\Pi$. 
Here, we develop the large-$q$ expansion of $\Pi(\mathbf q,i\nu_m)$ with finite $\nu_m$ and derive the asymptotic expression of the static XC kernel $\Kxc(q)$ in the large-$q$ limit.

\begin{figure}
    \centering
    \includegraphics[width=0.95\linewidth]{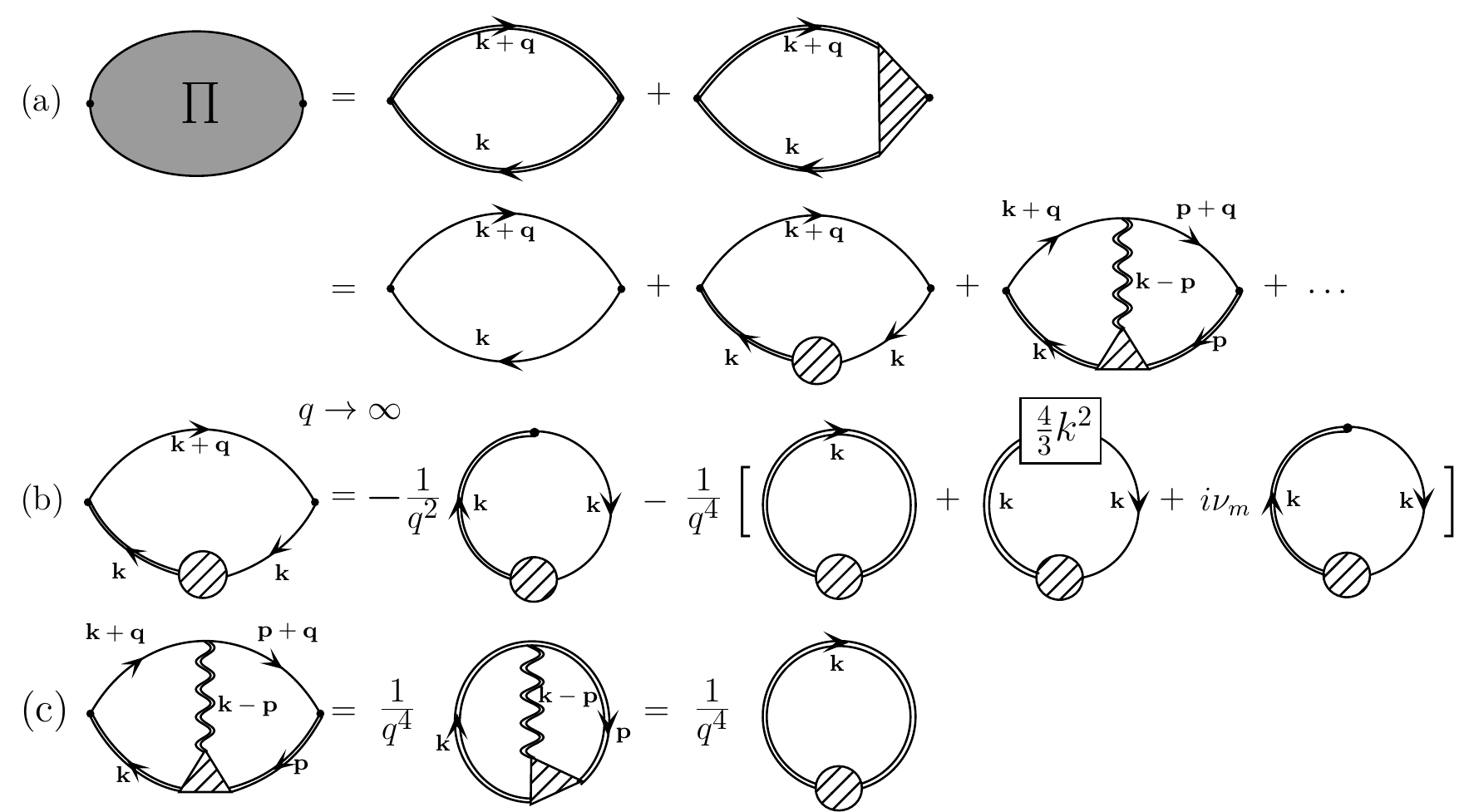}
    \caption{(a) The perturbative expansion for the polarization function $\Pi(\mathbf q,i\nu_m)$ formulated with the standard Feynman diagrams. The diagrams plotted in the second line include all diagrams whose leading term is proportional to $1/q^2$ or $1/q^4$ in the large-$q$ expansion of $\Pi$. The solid and double lines represent the bare propagator $\mathcal G_0$ and the bold propagator $\mathcal G$, respectively. The double wave represents the effective potential $W$, the shaded circle represents the self-energy $\Sigma$, and the shaded triangular represents the one-interaction-irreducible vertex function $\Gamma$. (b) The large-$q$ expansion up to $1/q^4$ for the high-order diagrams having self-energy insertion without vertex corrections. The $1/q^2$ term and the last $1/q^4$ term with prefactor $i\nu_m$ equal zero according to Dyson's equation and the density conservation. The second $1/q^4$ term equals $8/3$ times the interaction-induced shift of the kinetic energy $\delta \Ek$. (c) The large-$q$ expansion up to $1/q^4$ for the high-order diagrams with one vertex correction. It cancels with the first $1/q^4$ term in (b). The second equation is given by Hedin's equations. }
    \label{fig:diagrams}
\end{figure}

We first formulate the perturbative expansion for $\Pi(q,i\nu_m)$ by Hedin's equations. As shown in Fig.~\ref{fig:diagrams}(a), each contribution of $\Pi$ is computed by multiplying the amplitudes, including bare propagator $\mathcal{G}_0$, bold propagator $\mathcal G$, effective potential $W$, self-energy $\Sigma$, and vertex function $\Gamma$. Expanding $\mathcal{G}_0$ with flowing external momentum $q$ as
\begin{equation}
\begin{aligned}
&\mathcal{G}_0(\mathbf {k+q}, i\omega_n+i\nu_m) = \frac{1}{i\omega_n+i\nu_m -(\mathbf {k+q})^2 +\mu}\\
=& \frac{-1}{q^2} -\frac{\mathcal{G}_0^{-1}(\mathbf {k}, i\omega_n)-2\mathbf{k}\cdot \mathbf{q}+4(\mathbf{k}\cdot \mathbf{\hat q})^2 +i\nu_m}{q^4} +O(q^{-6})\,, 
\end{aligned}
\label{eq:expansion_G0}
\end{equation}
where $\omega_n=(2n+1)\pi /\beta$ is the fermionic Matsubara frequency and $\mu$ is the chemical potential, we find that each bare propagator $\mathcal{G}_0$ with flowing $q$ contributes $-1/q^2$ and the subleading $1/q^4$ terms for $\Pi$. Therefore, we can determine the dominant large-$q$ contribution of each polarization diagram as $\sim 1/q^{2\ell}$, where $\ell$ is the least number of the propagator lines with flowing $q$. As shown in the second line of Fig.~\ref{fig:diagrams}(a), we give all polarization diagrams with $\ell=1$ or $2$, including the zero-order diagram $\chi_0$, the high-order diagrams having self-energy in insertion without vertex corrections, and the high-order diagrams with only one vertex correction.

We next present the derivation of the large-$q$ expansion of $\chi_0$ for any finite Matsubara frequency $\nu_m$, which reads
\begin{equation}
    \begin{aligned}
    \chi_0(q, i\nu_m) &=  -\frac{2}{\beta}  \int d \mathbf{k}  \sum_{n} \mathcal{G}_0(\mathbf{k}, i\omega_n) \mathcal{G}_0(\mathbf{k}+\mathbf{q}, i\omega_n+i\nu_m) \\
&=-2 \int d \mathbf{k} \frac{f(\epsilon_{\mathbf{k}})-f(\epsilon_{\mathbf{k}+\mathbf{q}})}{i\nu_m+\epsilon_{\mathbf{k}}-\epsilon_{\mathbf{k}+\mathbf{q}}} \\
&=\int_0^{\infty} \frac{dk}{(2\pi)^2}\frac{k}{q} f(\epsilon_{\mathbf k}) \ln\frac{\nu_m^2 + (q^2+2kq)^2}{\nu_m^2+(q^2-2kq)^2} \,,
    \end{aligned}
\end{equation}
where the factor 2 comes from spin summation and $f(\epsilon_{\mathbf{k}})=\sum_n \mathcal G_0(\mathbf k, i\omega_n) e^{i\omega_n 0^+}$ is the momentum distribution of noninteracting fermions. Expanding the integrand for $q\to \infty$, we obtain 
\begin{equation}
    \begin{aligned}
     \chi_0(q, i\nu_m) &= 8\int \frac{dk}{(2\pi)^2}k^2 f(\epsilon_{\mathbf k})\left[\frac{1}{q^2}+\frac{4k^2}{3q^4} + \mathcal O(q^{-6}) \right] \\
     & = \frac{2\rho}{q^2} + \frac{8}{3} \frac{\Eko}{q^4} + \mathcal O(q^{-6}) \,.
    \end{aligned}
    \label{eq:expansion_chi0}
\end{equation}
Here, density $\rho \equiv (4\pi r_s^3/3)^{-1}$ is equal to $2\int d\mathbf k f(\epsilon_{\mathbf k})$ and the noninteracting kinetic energy $\Eko$ is equal to $2\int d\mathbf k k^2 f(\epsilon_{\mathbf k})$. 
Moreover, the large-$q$ expansion of $\chi_0$ can also be directly derived using the $\mathcal G_0$ expansion in Eq.\eqref{eq:expansion_G0}. 

We now turn to the large-$q$ expansion of the high-order diagrams up to $1/q^4$. For the diagrams in Fig.~\ref{fig:diagrams}(b), the $1/q^2$-term $\int d\mathbf{k} \sum_{n} \mathcal{G} \Sigma \mathcal{G}_0(\mathbf k,i\omega_n)$ can be rewritten as $\int d\mathbf{k} \sum_{n} \left[\mathcal{G} - \mathcal{G}_0\right](\mathbf k,i\omega_n)$ via Dyson's equation, which vanishes due to the density conservation. The $1/q^4$ terms have three parts: the first one comes from the $\mathcal G_0^{-1}/q^4$ contribution in Eq.\eqref{eq:expansion_G0}; the second one is obtained after integrating the angle in $4(\mathbf{k} \cdot \mathbf{\hat q})^2/q^4$ as
\begin{equation}
    \begin{aligned}
  &  2 \int d \mathbf k  \frac{4(\mathbf{k} \cdot \mathbf{\hat q})^2}{q^4} \sum_{n} \mathcal{G} \Sigma \mathcal{G}_0(\mathbf k,i\omega_n)  \\
 =& 2\int_0^{\infty} \frac{dk}{(2\pi)^2} \int_{-1}^{1} d\cos\theta \frac{4k^4\cos^2\theta}{q^4} \sum_{n} \mathcal{G} \Sigma \mathcal{G}_0(k,i\omega_n) \\
 =& 2 \int d\mathbf k \frac{4k^2}{3q^4} \sum_{n} \mathcal{G} \Sigma \mathcal{G}_0(k,i\omega_n) \,,
    \end{aligned}
    \label{eq:exp_highorder}
\end{equation}
where the factor 2 comes from spin summation; and the third one comes from the $i\nu_m/q^4$ contribution in Eq.~\eqref{eq:expansion_G0}, which vanishes like the $1/q^2$ term.
Similarly, we obtain the $1/q^4$ term of the diagrams with one vertex correction in Fig.~\ref{fig:diagrams}(c), which exactly cancels the first $1/q^4$ term in Fig.~\ref{fig:diagrams}(b). Consequently, the only existed term is the $1/q^4$ term in Eq.~\eqref{eq:exp_highorder}.
Finally, we obtain the large-$q$ expansion of high-order polarization as
\begin{equation}  
\begin{aligned}
& \lim_{q\to \infty} \left[ \Pi - \chi_0 \right] (q,i\nu_m) \\
=& 2 \int d\mathbf k \frac{8k^2}{3q^4}  \sum_{n} \mathcal{G} \Sigma \mathcal{G}_0(\mathbf k,i\omega_n) +O(q^{-6}) \\ 
=& \frac{16}{3q^4} \int d\mathbf{k} k^2 \sum_{n} \mathcal{G}(\mathbf k,i\omega_n) - \mathcal{G}_0(\mathbf k,i\omega_n) +O(q^{-6}) \\
=& \frac{8}{3q^4} (\Ek-\Eko) +O(q^{-6})
\,,
\end{aligned}
\label{eq:expansion_Pi}
\end{equation}
where the factor 2 is the symmetry factor.
Interestingly, the coefficient of the $1/q^4$ term is exactly 8/3 times the interaction-induced shift of the kinetic energy $\delta \Ek \equiv \Ek-\Eko$. It is noted that the leading-order term does not depend on $\nu_m$, which should exist on high-order terms.

Substituting Eq.\eqref{eq:expansion_chi0} and Eq.\eqref{eq:expansion_Pi} into Eq.\eqref{eq:GPi}, we obtain the asymptotic formula of $\Kxc(q)$ in the large-$q$ limit as
\begin{equation}
\begin{aligned}
    \lim_{q\to \infty} \Kxc(q;r_s, T) &= -\frac{2}{3\rho^2} \delta \Ek(r_s,T) + \mathcal O(q^{-2}) \\
    &= -\frac{32\pi^2 r_s^6 }{27} \delta \Ek(r_s,T) + \mathcal O(q^{-2}) \,.
\end{aligned}
\end{equation}
This formula holds for the dynamic XC kernel with any finite frequency since the frequency dependence does not exist in the leading order.

\section{Calculation of the kinetic energy}
The kinetic energy of the three-dimensional uniform electron gas reads
\begin{equation}
    \Ek  =  2\int d \mathbf k k^2 \mathcal G(\mathbf k, \tau=0^- ) \,,
\end{equation}
where $\tau$ is imaginary time.
By setting each component with $q$ in the polarization diagrams $\mathcal G(\mathbf k +\mathbf q, \tau)\mathcal G(\mathbf k, -\tau)$ to one, all the diagrams of the bold propagator $\mathcal G(\mathbf k, \tau)$ are generated using the Feynman diagrammatic expansion for the polarization in the variational scheme~\cite{chen_combined_2019}. Therefore, we can directly calculate $\Ek$ by the variational diagrammatic Monte Carlo (VDMC) method. Moreover, we calculate the noninteracting kinetic energy 
\begin{equation}
\begin{aligned}
    \Eko &= 2\int d \mathbf k k^2 \mathcal G_0(\mathbf k, \tau=0^- )  \\
    & = \int dk \frac{k^4}{\pi^2} \frac{1}{e^{k^2-\mu}+1} \,,
\end{aligned}
\end{equation}
by a one-dimensional numerical integration.
All the data of $\Ek$ and $\Eko$ are given in Tables~\ref{tab:Ek}.

\begin{table}[]
    \centering
 \begin{tabular}{r@{\hspace{5.5mm}} r@{\hspace{5.5mm}} r@{\hspace{5.5mm}} r@{\hspace{5.5mm}}}
 \hline
 $\theta$ & $E_{\rm k0}$ & $ E_{\rm k}$  
 & $\delta \Ek$ \\
 \hline
0.0625 & 0.536\,000  & 0.552\,14(142)  & 0.016\,14\\ 
  0.25 & 0.647\,785  & 0.653\,23(236)  & 0.005\,45 \\
  0.4  & 0.790\,910  & 0.789\,27(197)  & -0.001\,64\\
  0.5  & 0.898\,322  & 0.891\,78(188)  & -0.006\,54\\
    1  & 1.491\,929  & 1.473\,95(200)  & -0.017\,98\\
    2  & 2.761\,287  & 2.742\,35(34)  & -0.018\,94\\
    4  & 5.363\,294  & 5.349\,57(20)  & -0.013\,72\\
    6  & 7.985\,161  & 7.974\,66(20)  & -0.010\,50\\
    8  & 10.613\,473  & 10.605\,81(37)  & -0.007\,66\\
   12  & 15.877\,866  & 15.871\,72(17)  & -0.006\,15\\
   16  & 21.146\,840  & 21.141\,95(18)  &-0.004\,89\\
 \hline
    \end{tabular}
    \caption{Kinetic energies $E_{\rm k}$ from VDMC simulations and ideal kinetic energies $E_{\rm k0}$ from numerical integration (errors can be neglected compared to $E_{\rm k}$’s)   for $r_s=1$. }
    \label{tab:Ek}
\end{table}

\section{Data table of the local field correction}
All the local field correction data given in Tables~\ref{tab:LFC} and \ref{tab:LFC_rs2} are obtained by performing VDMC simulations for the truncation order $N=5$ and including extrapolation errors.  
\clearpage

\begin{center}
\tablefirsthead{%
\hline
 $\theta$ & $q/k_{\textsc f}$ & $G$  &error \\
 \hline}
 \tablehead{%
\hline
 $\theta$ & $q/k_{\textsc f}$ & $G$  &error \\
 \hline}
 \tablelasttail{\hline}
 \topcaption{\label{tab:LFC}Static local field corrections $G(q,\theta)$ for $r_s=1$ from VDMC simulations. }
\begin{supertabular}{r@{\hspace{5.5mm}} r@{\hspace{5.5mm}} c@{\hspace{5.5mm}} c@{\hspace{5.5mm}}}
0.0625 &0.1875 &0.0090 &0.0003  \\
0.0625 &0.3750 &0.0359 &0.0012  \\
0.0625 &0.5625 &0.0814 &0.0027  \\
0.0625 &0.7500 &0.1463 &0.0044  \\
0.0625 &0.9375 &0.2329 &0.0065  \\
0.0625 &1.1250 &0.3448 &0.0092  \\
0.0625 &1.3125 &0.4867 &0.0123  \\
0.0625 &1.5000 &0.6609 &0.0159  \\
0.0625 &1.6875 &0.8715 &0.0204  \\
0.0625 &1.8750 &1.0805 &0.0259  \\
0.0625 &2.0000 &1.1175 &0.0241  \\
0.0625 &2.2500 &1.0295 &0.0368  \\
0.0625 &2.5000 &1.0190 &0.0383  \\
0.0625 &2.7500 &1.0270 &0.0414  \\
0.0625 &3.0000 &1.0318 &0.0477  \\
0.0625 &3.5000 &1.0520 &0.0525  \\
0.0625 &4.0000 &1.1068 &0.0660  \\
0.0625 &4.5000 &1.1931 &0.0839  \\
0.0625 &5.0000 &1.2956 &0.1056  \\
0.0625 &5.5000 &1.4231 &0.1321  \\
0.0625 &6.0000 &1.5537 &0.1640  \\
0.0625 &6.5000 &1.6908 &0.1896  \\
0.0625 &7.0000 &1.7712 &0.3640  \\
0.0625 &8.0000 &2.0620 &0.4874  \\
0.0625 &9.0000 &2.3935 &0.5872  \\
0.0625 &10.0000 &2.7799 &0.7599 \\
\hline
0.25 &0.1875 &0.0093 &0.0003  \\
0.25 &0.3750 &0.0372 &0.0011  \\
0.25 &0.5625 &0.0842 &0.0021  \\
0.25 &0.7500 &0.1515 &0.0032  \\
0.25 &0.9375 &0.2406 &0.0045  \\
0.25 &1.1250 &0.3537 &0.0059  \\
0.25 &1.3125 &0.4914 &0.0074  \\
0.25 &1.5000 &0.6510 &0.0090  \\
0.25 &1.6875 &0.8205 &0.0111  \\
0.25 &1.8750 &0.9710 &0.0144  \\
0.25 &2.0625 &1.0581 &0.0181  \\
0.25 &2.2500 &1.0599 &0.0275  \\
0.25 &2.4375 &1.0019 &0.0409  \\
0.25 &2.6250 &0.9356 &0.0558  \\
0.25 &2.8125 &0.9053 &0.0731  \\
0.25 &3.0000 &0.8317 &0.0595  \\
0.25 &4.0000 &0.8502 &0.0800  \\
0.25 &5.0000 &0.9103 &0.1311  \\
0.25 &6.0000 &1.0118 &0.2118  \\
0.25 &7.0000 &1.1298 &0.3172  \\
0.25 &8.0000 &1.2461 &0.4547  \\
 \hline
0.4 &0.2500 &0.0176 &0.0005  \\
0.4 &0.5000 &0.0705 &0.0015  \\
0.4 &0.7500 &0.1597 &0.0028  \\
0.4 &1.0000 &0.2862 &0.0042  \\
0.4 &1.2500 &0.4497 &0.0062  \\
0.4 &1.5000 &0.6415 &0.0080  \\
0.4 &1.7500 &0.8389 &0.0103  \\
0.4 &2.0000 &0.9970 &0.0130  \\
0.4 &2.2500 &1.0589 &0.0175  \\
0.4 &2.5000 &1.0041 &0.0272  \\
0.4 &2.7500 &0.8297 &0.0396  \\
0.4 &3.0000 &0.6830 &0.0428  \\
0.4 &3.2500 &0.5817 &0.0462  \\
0.4 &3.5000 &0.5303 &0.0503  \\
0.4 &3.7500 &0.5106 &0.0559  \\
0.4 &4.0000 &0.4992 &0.0622  \\
0.4 &4.5000 &0.4808 &0.0791  \\
0.4 &5.0000 &0.4529 &0.1021  \\
0.4 &5.5000 &0.4190 &0.1284  \\
0.4 &6.0000 &0.3767 &0.1529  \\
0.4 &6.5000 &0.3386 &0.1890  \\
0.4 &7.0000 &0.3027 &0.2293  \\
0.4 &7.5000 &0.2442 &0.2692  \\
 \hline
1 &0.2500 &0.0196 &0.0004  \\
1 &0.5000 &0.0766 &0.0015  \\
1 &0.7500 &0.1670 &0.0031  \\
1 &1.0000 &0.2849 &0.0048  \\
1 &1.2500 &0.4239 &0.0061  \\
1 &1.5000 &0.5762 &0.0072  \\
1 &1.7500 &0.7304 &0.0071  \\
1 &2.0000 &0.8732 &0.0070  \\
1 &2.2500 &0.9884 &0.0055  \\
1 &2.5000 &1.0570 &0.0063  \\
1 &2.7500 &1.0607 &0.0045  \\
1 &3.0000 &0.9884 &0.0101  \\
1 &3.2500 &0.8481 &0.0107  \\
1 &3.5000 &0.6317 &0.0270  \\
1 &3.7500 &0.3957 &0.0302  \\
1 &4.0000 &0.0936 &0.0613  \\
1 &4.5000 &-0.3062 &0.0562 \\  
1 &5.0000 &-0.6051 &0.1124 \\
1 &5.5000 &-0.7602 &0.0971 \\
1 &6.5000 &-1.0423 &0.1331 \\
1 &7.5000 &-1.4262 &0.1711 \\
1 &8.5000 &-1.9074 &0.2438 \\
 \hline
2 &0.5000 &0.0669 &0.0025  \\
2 &1.0000 &0.2343 &0.0061  \\
2 &1.5000 &0.4562 &0.0078  \\
2 &2.0000 &0.6968 &0.0078  \\
2 &2.5000 &0.9168 &0.0076  \\
2 &3.0000 &1.0654 &0.0039  \\
2 &3.5000 &1.0821 &0.0039  \\
2 &4.0000 &0.9199 &0.0066  \\
2 &4.5000 &0.5668 &0.0125  \\
2 &5.0000 &0.0631 &0.0228  \\
2 &6.0000 &-1.0253 &0.0531  \\
2 &7.0000 &-1.7685 &0.0918  \\
2 &8.0000 &-2.2129 &0.1404  \\
2 &9.0000 &-2.6287 &0.2060  \\
 \hline
4 &0.3125 &0.0197 &0.0015  \\
4 &0.6250 &0.0720 &0.0044  \\
4 &0.9375 &0.1452 &0.0067  \\
4 &1.2500 &0.2338 &0.0081  \\
4 &1.5625 &0.3297 &0.0090  \\
4 &1.8750 &0.4382 &0.0096  \\
4 &2.1875 &0.5505 &0.0106  \\
4 &2.5000 &0.6621 &0.0070  \\
4 &2.8125 &0.7729 &0.0068  \\
4 &3.1250 &0.8770 &0.0065  \\
4 &3.4375 &0.9690 &0.0063  \\
4 &3.7500 &1.0434 &0.0062  \\
4 &4.0625 &1.0941 &0.0062  \\
4 &4.3750 &1.1146 &0.0064  \\
4 &4.6875 &1.0996 &0.0067  \\
4 &5.0000 &1.0438 &0.0106  \\
4 &6.0000 &0.5620 &0.0183  \\
4 &7.0000 &-0.3278 &0.0333 \\
4 &8.0000 &-1.3619 &0.0617 \\
4 &9.0000 &-2.2441 &0.1022 \\
4 &10.0000 &-2.8511 &0.1170  \\
4 &11.0000 &-3.3196 &0.2457  \\
4 &12.0000 &-3.5672 &0.3957  \\
4 &13.0000 &-4.1618 &0.4452  \\
4 &14.0000 &-4.6946 &0.4876  \\
 \hline
8 &0.3125 &0.0123 &0.0013  \\
8 &0.6250 &0.0443 &0.0039  \\
8 &0.9375 &0.0883 &0.0059  \\
8 &1.2500 &0.1404 &0.0075  \\
8 &1.5625 &0.1986 &0.0090  \\
8 &1.8750 &0.2639 &0.0107  \\
8 &2.1875 &0.3373 &0.0121  \\
8 &2.5000 &0.4082 &0.0146  \\
8 &2.8125 &0.4841 &0.0174  \\
8 &3.1250 &0.5677 &0.0215  \\
8 &3.4375 &0.6555 &0.0248  \\
8 &3.7500 &0.7297 &0.0315  \\
8 &4.0625 &0.7957 &0.0397  \\
8 &4.3750 &0.8760 &0.0505  \\
8 &4.6875 &0.9496 &0.0637  \\
8 &5.0000 &1.0204 &0.0128  \\
8 &6.0000 &1.1164 &0.0176  \\
8 &7.0000 &1.0156 &0.0252  \\
8 &8.0000 &0.6601 &0.0436  \\
8 &9.0000 &0.0443 &0.0707  \\
8 &10.0000 &-0.7639 &0.1337  \\
8 &11.0000 &-1.6473 &0.2022  \\
8 &12.0000 &-2.4692 &0.3048  \\
    \end{supertabular}
\end{center}

\begin{table}[]
    \centering
 \begin{tabular}{r@{\hspace{5.5mm}} r@{\hspace{5.5mm}} c@{\hspace{5.5mm}} c@{\hspace{5.5mm}}}
 \hline
 $\theta$ & $q/k_{\textsc f}$ & $G$  &error \\
 \hline
4 &0.5000 &0.0537 &0.0052  \\
4 &1.0000 &0.1869 &0.0138  \\
4 &1.5000 &0.3577 &0.0185  \\
4 &2.0000 &0.5424 &0.0190  \\
4 &2.5000 &0.7277 &0.0171  \\
4 &3.0000 &0.9006 &0.0143  \\
4 &3.5000 &1.0460 &0.0119  \\
4 &4.0000 &1.1457 &0.0103  \\
4 &4.5000 &1.1808 &0.0107  \\
4 &5.0000 &1.1340 &0.0131  \\
4 &5.5000 &0.9928 &0.0188  \\
4 &6.0000 &0.7566 &0.0274  \\
4 &6.5000 &0.4359 &0.0399  \\
4 &7.0000 &0.0569 &0.0557  \\
4 &7.5000 &-0.3491 &0.0720   \\
4 &8.0000 &-0.7437 &0.0987   \\
4 &8.5000 &-1.1015 &0.1266   \\
4 &9.0000 &-1.4006 &0.1485   \\
4 &9.5000 &-1.6456 &0.1853   \\
4 &10.0000 &-1.8453 &0.1878  \\
4 &10.5000 &-2.0122 &0.2696  \\
4 &11.0000 &-2.1482 &0.3101  \\
4 &11.5000 &-2.3023 &0.3541  \\
\hline
8 &0.5000 &0.0349 &0.0041  \\
8 &1.0000 &0.1204 &0.0106  \\
8 &1.5000 &0.2293 &0.0140  \\
8 &2.0000 &0.3497 &0.0147  \\
8 &2.5000 &0.4774 &0.0142  \\
8 &3.0000 &0.6088 &0.0134  \\
8 &3.5000 &0.7397 &0.0126  \\
8 &4.0000 &0.8654 &0.0122  \\
8 &4.5000 &0.9789 &0.0121  \\
8 &5.0000 &1.0731 &0.0122  \\
8 &5.5000 &1.1398 &0.0131  \\
8 &6.0000 &1.1703 &0.0144  \\
8 &6.5000 &1.1573 &0.0159  \\
8 &7.0000 &1.0937 &0.0183  \\
8 &7.5000 &0.9738 &0.0215  \\
8 &8.0000 &0.7961 &0.0291  \\
8 &8.5000 &0.5616 &0.0343  \\
8 &9.0000 &0.2772 &0.0536  \\
8 &9.5000 &-0.0498 &0.0514 \\ 
8 &10.0000 &-0.4057 &0.0524 \\ 
8 &10.5000 &-0.7733 &0.0865 \\
8 &11.0000 &-1.1386 &0.1455 \\
8 &11.5000 &-1.4967 &0.1431 \\
\hline
    \end{tabular}
    \caption{Static local field corrections $G(q,\theta)$ for $r_s=2$ from VDMC simulations. }
    \label{tab:LFC_rs2}
\end{table}

\bibliographystyle{apsrev}
\bibliography{references}